\documentclass[runningheads,orivec]{llncs}
\usepackage{amsmath}
\usepackage{tikz}
\usepackage{graphicx}
\usepackage{mdframed}
\newmdenv[linewidth=0.5pt,roundcorner=0pt,innerleftmargin=6pt,innerrightmargin=6pt,innertopmargin=5pt,innerbottommargin=5pt,skipabove=0pt,skipbelow=4pt]{figureframe}
\usepackage{xcolor}
\usepackage{fancyvrb}
\DefineVerbatimEnvironment{code}{Verbatim}{%
  fontsize=\footnotesize,
  baselinestretch=0.90,
  xleftmargin=1.5em
}
\DefineVerbatimEnvironment{session}{Verbatim}{%
  fontsize=\footnotesize,
  baselinestretch=0.80,
  xleftmargin=2.5em
}
\DefineVerbatimEnvironment{sessionwide}{Verbatim}{%
  fontsize=\scriptsize,
  baselinestretch=0.80,
  xleftmargin=2.5em
}

\newcommand{\Language}{Scratchpad II}
\DeclareMathOperator{\ldiv}{ldiv}
\DeclareMathOperator{\rdiv}{rdiv}
\DeclareMathOperator{\ldquo}{ldquo}
\DeclareMathOperator{\rdquo}{rdquo}
\DeclareMathOperator{\ldrem}{ldrem}
\DeclareMathOperator{\rdrem}{rdrem}
\DeclareMathOperator{\lgcd}{lgcd}
\DeclareMathOperator{\rgcd}{rgcd}
\DeclareMathOperator{\llcm}{llcm}
\DeclareMathOperator{\rlcm}{rlcm}
\DeclareMathOperator{\degree}{degree}
\DeclareMathOperator{\lc}{lc}
\newcommand{\return}{\text{return}}
\title{\mbox{Algorithms for} \mbox{Linear Ordinary Differential Operators}%
\thanks{This paper was accepted for the
\textit{Conference on Computers and Mathematics}, held at Stanford,
California, 30 July--1 August 1986. The proceedings, however, did not
appear. This is a translation of the contemporaneous Script/VM to LaTeX.}}
\titlerunning{Algorithms for Linear Ordinary Differential Operators}
\author{Jean Della Dora \and Stephen M. Watt}
\authorrunning{J. Della Dora and S. M. Watt}
\institute{Institut IMAG, Laboratoire TIM3, BP 68,\\
38402 Saint Martin d'H\`eres Cedex, France
\and
IBM Thomas J. Watson Research Center,\\
Yorktown Heights, New York 10598, USA}

\begin{document}
\maketitle

\begin{abstract}
We describe an implementation in Scratchpad II of linear ordinary
 differential operators over a differential ring, acting on a module
 equipped with a compatible derivation.  The abstract data type
 facilities of the system allow such operators to be represented and
 manipulated as first-class objects while retaining the usual notation
 for operator application.  For coefficients in a field, we give
 constructive algorithms for left and right division, greatest common
 divisors, least common multiples, and an extended Euclidean algorithm;
 the right-hand constructions may be obtained from the corresponding
 left-hand constructions in the opposite ring.  We also discuss
 pseudo-division over polynomial coefficient rings and an Ore
 localization yielding a right field of fractions.  Finally, we apply
 this operator arithmetic to factorization of ordinary differential
 equations, using the associated Riccati equation and Newton polygons
 to analyze possible singular parts of factors.  Examples include
 operators with constant, elementary-function, rational-function, and
 matrix coefficients.
\end{abstract}

\section{Introduction}

We present an implementation of differential operators.
LODO(A, M) is the domain of linear ordinary differential operators
over an A-module M, for a differential ring A.
We first describe how Scratchpad II allows users to create
operators as first class objects and show how these mechanisms
are used in the case of our differential operator domain.
The main body of the paper examines the constructive algebraic aspects
of these operators.  In particular, when A is a field we have algorithms
for left and right division of operators, left and right GCDs, left and
right LCMs, an extended Euclidean algorithm, and
a localization process for constructing the right field of
fractions.
We also examine the problem of operator factorization
for the solution of ordinary differential equations.

\section{Linear Ordinary Differential Operators}
\subsection{Operators in \Language{}}

Although operator arithmetic is one of the more useful tools of the
applied mathematician, one of the long-standing shortcomings of
existing computer algebra systems is their lack of support for
operator algebra.
As far as we are aware, operator algebra has not been part of the initial
design of any computer algebra system.
After-the-fact addition of classes of operators can
be awkward, requiring either special functions for operator manipulation
(``opPlus'', ``opTimes'', etc.), which is ugly,
or requiring modification of the base system, which is impossible
for most users.

Here, we show how the
abstract data type architecture of \Language{} allows for  simple
creation of user defined operators.

\Language{} addresses the extensibility issue by providing generic
operations on abstract data types
and by making application itself an operation.
For example, the domain \texttt{A $\rightarrow$ B} of
functions from A to B exports the application operation

\begin{center}
``.'': (A $\rightarrow$ B, A) $\rightarrow$ B
\end{center}
\noindent
which takes a mapping from A to B and an
element of A to produce an element of B.

The notion of generic application is fundamental to the system and
is used not only for function invocation but also
for subscripting arrays, list element extraction, record field
selection and indexing tables.
Two syntaxes are provided for application, one associating
to the left and another to the right:

\begin{center}
f a b   means  f applied to (a applied to b) \\
f.a.b   means  (f applied to a) applied to b
\end{center}

To create a ring R of operators on a set S in \Language{},
one adds the operation

\begin{center}
``.'': (R, S) $\rightarrow$ S
\end{center}
\noindent
to the constructor of R.
\begin{figure}[t]
\begin{figureframe}
\begin{small}
\begin{code}
LinearOrdinaryDifferentialOperator(A, M): DOcategory == DOcapsule 
  where
    NNI ==> NonNegativeInteger
    SUP ==> SparseUnivariatePolynomial
 
    A:  DifferentialRing
    M:  Module(A) with deriv: $ -> $
 
    DOcategory == GeneralPolyWithoutCommutativity(A,NNI) with
       D:      () -> $
       ".":    ($, M) -> M
 
       if A has commutative("*") and A has constant(deriv) then
          commutative("*")
 
    DOcapsule == SUP(A) add
       Rep := SUP(A)
 
       a, b: $
       u: M
 
       coerce(a): Expression == .....
       D() == .....
       a.u == .....
       a*b == .....
\end{code}

\end{small}
\end{figureframe}
\caption{A Skeleton for LODO}
\label{fig-skel}
\end{figure}

\subsection{The Domain LODO(A,M)}

Using this facility, we have defined a domain of linear ordinary
differential operators in a general context.
We decided to restrict our attention initially to this
domain for a number of reasons.
First, this domain is fairly well understood and has some nice
algebraic properties.
Second, we felt that if we could create the differential operator
domain using only user-level facilities, then it could be used as an
example for other operator domains.
Finally, linear ordinary differential operators happened to be exactly
what we needed for some
work we were doing on ordinary differential equations.

Linear ordinary differential operators (LODO, for short) can be thought
of as non-commutative polynomials in \(\partial\), where \(\partial\)
is the derivation for the set upon which they act.
Let us fix \(A\) as the coefficient ring and \(M\) as the set
upon which the operators act.
Operator multiplication is defined by
\[
(f * g) ~ \phi = f ( g (\phi)).
\]
so \(\partial\) also acts as a compatible derivation on the
coefficient ring.
Application of a degree zero operator is simply a multiplication
\begin{center}
``*'': (A, M) $\rightarrow$ M
\end{center}
so \(M\) must be an \(A\)-Module with a derivation, at least.

With this in mind, let us define
\texttt{LODO}(A,M) as the domain
of linear ordinary differential operators over
an A-module M (with derivation), for a differential ring A.
This includes the
cases of operators which are polynomials in \(\partial\) acting on
scalars or vectors depending on a single variable.
The coefficients of the
operator polynomials can be integers, rational functions, matrices
or elements of other domains.

It is natural to represent the operators internally as polynomials,
and this is what we do.   \Language{}'s data abstraction mechanisms make
the details of the implementation completely inaccessible outside
the data type constructor.  It is therefore completely safe
to implement a non-commutative data type (\texttt{LODO(A, M)})
using a commutative type (\texttt{SparseUnivariatePolynomial(A)})
so long as care is taken to export the new multiplication function
and not to use the SparseUnivariatePolynomial multiplication within
the implementation.

Figure~\ref{fig-skel}
shows an outline of a basic definition for LODO.  The
elided implementations are discussed in the next section,
along with discussion of additional functions.

The remainder of this section gives examples of the use of the LODO
domain.
\subsection{Example: Differential operators with constant coefficients}

We begin by making some type assignments and declaring \textit{Dx}
to be a linear differential operator.
\begin{session}
PZ := UPDR(P[x]I, I);
 
PQ := UPDR(P[x]RN,RN);
 
Dx: LODO(RN, PQ)
\end{session}

\texttt{PZ} is the domain of univariate polynomials in \(x\)
over the integers with some additional properties that make it into
a differential ring.
Similarly, \texttt{PQ} is a domain of
univariate polynomials over the rationals.
(Within a few months one should be able to use \texttt{P[x] I}
and \texttt{P[x] RN} directly, without resorting to the
\texttt{UPDR} constructor.)

Operators are created as polynomials in \textit{D()}.
\begin{small}
\begin{session}
Dx := D()
 
   (5)  "D"
\end{session}
\begin{session}
a  := Dx  + 1
 
   (6)  D + 1
\end{session}
\begin{session}
b  := a + 1/2*Dx**2 - 1/2
 
          1   2        1
   (7)  (---)D  + D + ---
          2            2
\end{session}
\begin{session}
c := (1/9)*b*(a + b)**2
 
   (8)
        1    6     5    5     13   4     19   3    79   2     7        1
      (----)D  + (----)D  + (----)D  + (----)D + (----)D  + (----)D + ---
        72         36         24         18        72         12       8
\end{session}

\end{small}
\noindent
To apply the operator \(a\) to the value \(p\) the usual
function call syntax is used.
\begin{small}
\begin{session}
p: PQ := 4*x**2 + 2/3;
 
a p
 
           2         2
   (10)  4x  + 8x + ---
                     3
\end{session}

\end{small}
\noindent
Operator multiplication is defined by the identity
\( (a * b) p = a ( b ( p ) )\).
\begin{small}
\begin{session}
(a*b) p = a b p
 
           2          37      2          37
   (11)  2x  + 12x + ---- = 2x  + 12x + ----
                      3                  3
\end{session}

\end{small}
\noindent
Operator expressions may be applied directly.
\begin{small}
\begin{session}
(a**2 - 3/4*b + c) (p + 1)
 
           2     44       541
   (12)  3x  + (----)x + -----
                 3        36
\end{session}

\end{small}
\noindent
When the operator coefficients are rational, it is possible to factor.
\begin{small}
\begin{session}
factor c
 
           1          2       4
   (13)  (----)(D + 3) (D + 1)
           72
\end{session}

\end{small}

\subsection{Example: Differential operators with elementary function coefficients}

\noindent\textbf{Problem: }
find the first few coefficients of \({e ^ x} {x ^ {-i}}\)
in \(L _ 3 ~ \phi\), where
\[
L _ 3 = D ^ 3 + \frac{G}{x^2} D + \frac{H}{x^3} - 1
\]
and
\[
\phi = \sum _ {i = 0} ^ \infty 
\frac{s_i e^x}{x^i}
\]

We will compute the first five coefficients.
We start by defining the domain \texttt{EDF} to
be the elementary functions
viewed as a differential ring in the variable \(x\).
\begin{small}
\begin{session}
-- First we clear the values of some of the variables.
)clear value n Dx L3 phi rho num xnum
 
n := 5
 
   (32)  5
\end{session}
\begin{session}
EDF := ODR(SE, EF P I, x);
\end{session}

\end{small}
\noindent
Next we assign the differential operator \(L\) and the first
terms of the function \(\phi\).
\begin{small}
\begin{session}
(Dx, L3): LODO(EDF, EDF)
\end{session}
\begin{session}
Dx := D()
 
   (35)  "D"
\end{session}
\begin{session}
L3 := Dx**3 + G/x**2*Dx + H/x**3 - 1
 
                              3
          3     G        H - x
   (36)  D  + (----)D + --------
                 2          3
                x          x
\end{session}
\begin{session}
phi := +/[ s[i]*exp(x)/x**i for i in 0..n ]
 
                      2     3     4     5
          (s  +x*s  +x s  +x s  +x s  +x s )exp(x)
            5     4     3     2     1     0
   (37)  ------------------------------------------
                             5
                            x
\end{session}
\begin{session}
rho: EF P I := L3 phi
 
   (38)
                                   2
       ( (H + (x - 5)G - 15x  + 90x - 210)s
                                           5
                  2             3      2
      +  (x*H + (x  - 4x)G - 12x  + 60x  - 120x)s
                                                 4
           2      3     2       4      3      2
      +  (x H + (x  - 3x )G - 9x  + 36x  - 60x )s
                                                 3
           3      4     3       5      4      3
      +  (x H + (x  - 2x )G - 6x  + 18x  - 24x )s
                                                 2
           4      5    4       6     5     4
      +  (x H + (x  - x )G - 3x  + 6x  - 6x )s
                                              1
           5     6
      +  (x H + x G)s  )
                     0
               8
   * exp(x) / x
\end{session}

\end{small}
\noindent
We extract the numerator and remove the common factor of \(\exp(x)\).
\begin{small}
\begin{session}
-- ending the next line with ";" suppresses output
 
num: P I := rho / exp(x) * x**(n+3);
\end{session}

\end{small}
\noindent
To collect terms with like powers, make \(x\) the main variable.
\begin{small}
\begin{session}
xnum: P[x] P I := num;
\end{session}

\end{small}
\noindent
The answer is obtained by reading off the coefficients
corresponding to the powers
\(1/x^0\) to \(1/x^n\).
\begin{small}
\begin{sessionwide}
[ coef(xnum, n+3 - i) for i in 0..n ]
 
   (41)
 
   [0, 0, - 3s  + s G, - 6s  + s G + 6s  + s H,
              1    0       2    1      1    0
    - 9s  + s G + 18s  + s H - s G - 6s , - 12s  + s G + 36s  + s H - 2s G - 24s ]
        3    2       2    1     1      1       4    3       3    2      2       2
\end{sessionwide}

\end{small}
\subsection{Example: Differential operators with matrix coefficients acting on vectors}

Define the differential ring of 3 by 3 matrices of polynomials
in \(x\) over
the integers and the three dimensional vector space of these
polynomials.
\begin{small}
\begin{session}
)clear value Dx a b t p0 p1 p2 p q m
 
Mat  := SMDR(3, PZ);
 
Vect := DPMM(3, PZ, Mat, PZ);
\end{session}

\end{small}
\noindent
The matrix \(m\) will be used as a coefficient and
the vectors \(p\) and \(q\) will be operated upon.
\begin{small}
\begin{session}
t: PZ := x**2;
 
m := zero(3,3)$Mat;
m(0,0) := t;
m(0,1) := 1;
m(1,0) := 1;
m(1,1) := t**2;
m(2,2) := 4*t;
 
m
 
         [ 2         ]
         [x   1    0 ]
         [           ]
   (52)  [     4     ]
         [1   x    0 ]
         [           ]
         [          2]
         [0   0   4x ]

p0: PZ := 3*x**2 + 1;
p1: PZ := 2*x;
p2: PZ := 7*x**3 + 2*x;
p: Vect := [p0, p1, p2]$V(PZ)
 
            2          3
   (56)  [3x  + 1,2x,7x  + 2x]

q: Vect := m * p
 
            4    2        5     2        5     3
   (57)  [3x  + x  + 2x,2x  + 3x  + 1,28x  + 8x ]
\end{session}

\end{small}
The operation
\begin{center}
``*'': (\texttt{Mat}, \texttt{Vect}) $\rightarrow$ \texttt{Vect}
\end{center}
is defined so it makes sense to
define operators with coefficients in \textit{Mat} acting on
members of \textit{Vect}
\begin{small}
\begin{session}
(Dx, a, b): LODO(Mat, Vect)
\end{session}
\begin{session}
Dx := D()
 
   (59)  "D"
\end{session}
\begin{session}
a := Dx  + m
 
             [ 2         ]
             [x   1    0 ]
             [           ]
   (60)  D + [     4     ]
             [1   x    0 ]
             [           ]
             [          2]
             [0   0   4x ]
\end{session}
\begin{session}
b := m*Dx  + 1
 
         [ 2         ]
         [x   1    0 ]    [1  0  0]
         [           ]    [       ]
   (61)  [     4     ]D + [0  1  0]
         [1   x    0 ]    [       ]
         [           ]    [0  0  1]
         [          2]
         [0   0   4x ]
\end{session}
\begin{session}
a * b
 
   (62)
      [ 2         ]    [ 4              4    2                  ]    [ 2         ]
      [x   1    0 ]    [x  + 2x + 2    x  + x            0      ]    [x   1    0 ]
      [           ] 2  [                                        ]    [           ]
      [     4     ]D + [   4    2     8     3                   ]D + [     4     ]
      [1   x    0 ]    [  x  + x     x  + 4x  + 2        0      ]    [1   x    0 ]
      [           ]    [                                        ]    [           ]
      [          2]    [                              4         ]    [          2]
      [0   0   4x ]    [     0            0        16x  + 8x + 1]    [0   0   4x ]
\end{session}
\begin{session}
(a+b) (p + q)
 
   (63)
 
      6      5      4      3      2
   [3x  + 14x  + 17x  + 22x  + 10x  + 18x + 6,
 
      9      8     6      5      4      3     2
    2x  + 10x  + 3x  + 10x  + 16x  + 12x  + 7x  + 18x + 6,   

        7       6      5       4      3      2
    112x  + 560x  + 88x  + 320x  + 23x  + 53x  + 2x + 2]
\end{session}

\end{small}
\section{Operations on Linear Ordinary Differential Operators}
\subsection{Fundamental Operations}

The algorithms in this paper are described in the particular terms
of differential operators.  It should be noted, though, that
many of the algorithms apply to non-commutative polynomials in general.

We begin by describing the fundamental operations on ordinary
differential operators.

The members of \texttt{LODO(A,M)} are represented as
sparse univariate polynomials with coefficients in A.
The basic definition of LODO
must include the trivial operations of addition and subtraction.
These are inherited from \texttt{SUP(A)}.
Coerce to expression produces an output form with the
anonymous polynomial variable displayed as ``D''.
The function \textit{D}, which creates the operator \(\partial\)
is a simple monomial.
\vbox{
\begin{small}
\begin{code}
-- $   is LODO(A, M)
-- Rep is SUP(A)
 
a: $
u: M
 
coerce(a): Expression == output(a, "D":: Expression)$Rep
 
D() == monom(1,1)
\end{code}
\end{small}
}

Operator application is also quite straightforward
(see Figure~\ref{fig-appl}).
\begin{figure}[t]
\begin{figureframe}
\begin{small}
\begin{code}
a.u ==
    u, w, uderiv: M
    w := 0
    for i in 0..degree a repeat
        -- uderiv is the i-th derivative of u using deriv: M -> M
        uderiv := if i = 0 then u else deriv uderiv
        w := w + coef(a, i)*uderiv
    w
\end{code}

\end{small}
\end{figureframe}
\caption{Operator Application}
\label{fig-appl}
\end{figure}
The first non-trivial construct is the multiplication of two
members of LODO.
Multiplication is based on the following Heisenberg
commutation rule
\[
             \partial * x = x * \partial + 1
\]
and the repeated use of the Leibnitz rule.
Of course, the important point to note is the non-commutativity
of this operation.
The program for the multiplication is shown in
Figure~\ref{fig-mult}.
\begin{figure}[t]
\begin{figureframe}
\begin{small}
\begin{code}
a, b: $
 
a * b ==
    alpha, beta, bderiv: A
    aa, bb, r: $
    degaa, degbb, n: NonNegativeInteger
    aCi : Integer
 
    r  := 0
    aa := a
    while aa^=0 repeat
        degaa := degree aa
        alpha := lc aa
        aa    := red aa
 
        bb := b
        while bb^=0 repeat
            degbb := degree bb
            beta  := lc bb
            bb    := red bb
 
            for i in 0..degaa repeat
                if i=0 then
                    bderiv := beta   -- i-th deriv of beta
                    aCi    := 1      -- degaa choose i
                else
                    bderiv := deriv bderiv
                    aCi    := aCi * (degaa - i + 1) quo i
                d := degaa + degbb - i
                r := r + monom(d, aCi*alpha*bderiv)
    r
\end{code}

\end{small}
\end{figureframe}
\caption{Operator Multiplication}
\label{fig-mult}
\end{figure}
\subsection{Division}

When the coefficient ring is a field we can
define division operations.
Because the differential operators are non-commutative there are two
divisions, from the left and from the right, defined by
\begin{gather*}
{\ldiv(a,b) = [q,r] ~~~~ \text{iff} ~~~~ a=b*q+r,~~ \degree (r) < \degree (b) }
  \\
{\rdiv(a,b) = [s,t] ~~~~ \text{iff} ~~~~ a=s*b+t,~~ \degree (t) < \degree (b) }
\end{gather*}
In the literature, \(q\) is sometimes called the
\textit{left quotient}  and sometimes the
\textit{right quotient}.
For clarity, we call \(q\) the \textit{left division quotient}
and \(r\) the \textit{left division remainder}.
Likewise \(s\) and \(t\) are called the \textit{right division quotient} and \textit{remainder}.

The division algorithms are based on repeated subtraction.
We illustrate this with the code for left division in
Figure~\ref{fig-div}.
\begin{figure}[t]
\begin{figureframe}
\begin{small}
\begin{code}
-- ldiv(a,b)=[q,r]  means a=b*q+r
 
ldiv: ($, $) -> Record(quotient: $, remainder: $)
 
ldiv(a, b) ==
    q: $ := 0
    r: $ := a
    iv:A := inv lc b
    while degree r >= degree b and r ^= 0 repeat
        h := monom(degree r - degree b, iv*lc r)
        r := r - b*h
        q := q + h
    [q,r]
 
rdiv(a, b) ==
    q: $ := 0
    r: $ := a
    iv:A := inv lc b
    while degree r >= degree b and r ^= 0 repeat
        h := monom(degree r - degree b, iv*lc r)
        r := r - h*b
        q := q + h
    [q,r]
\end{code}

\end{small}
\end{figureframe}
\caption{Left and Right Division}
\label{fig-div}
\end{figure}

We sometimes want just the quotient or just the remainder.
Given the algorithms for left and right division,
the following operations can be implemented easily
\begin{gather*}
{{\ldquo(a, b) = q} ~~~~~~~ {\ldrem(a, b) = r}}
\\
{{\rdquo(a, b) = s} ~~~~~~~ {\rdrem(a, b) = t}}
\end{gather*}
for left (right) division quotient (remainder).

Having the preceding operations it makes sense to begin the arithmetical
part of a package for linear ordinary differential operators with
coefficients in the field \(K(x)\).
\subsection{Greatest Common Divisors}

Using the division operations, it is possible to compute GCDs via remainder
sequences.
Again, because of the non-commutativity of operator multiplication,
there are two GCDs, defined as follows:
\[
\begin{aligned}
lg=\lgcd(a,b)\quad\text{iff}\quad
 &lg\text{ has maximum degree such that}\\
 &\exists A,B,\quad a=lg*A\quad\text{and}\quad b=lg*B;
\end{aligned}
\]
\[
\begin{aligned}
rg=\rgcd(a,b)\quad\text{iff}\quad
 &rg\text{ has maximum degree such that}\\
 &\exists A,B,\quad a=A*rg\quad\text{and}\quad b=B*rg.
\end{aligned}
\]
\noindent
Both of the GCDs are computed using a Euclidean algorithm.
We show the algorithm for computing the right GCD in Figure~\ref{fig-gcd}.
Since the coefficients come from a field, it is possible to further demand that the GCDs be monic.
\begin{figure}[t]
\begin{figureframe}
\begin{small}
\begin{code}
lgcd(a,b) ==
    a = 0 =>b
    b = 0 =>a
    while degree b > 0 repeat
        (a,b) := (b, ldrem(a,b))
    if b=0 then a else b
 
rgcd(a,b) ==
    a = 0 =>b
    b = 0 =>a
    while degree b > 0 repeat
        (a,b) := (b, rdrem(a,b))
    if b=0 then a else b
\end{code}

\end{small}
\end{figureframe}
\caption{Left and Right GCD}
\label{fig-gcd}
\end{figure}
\subsection{Least Common Multiples}

Using each of the GCD operations it is possible to construct an LCM.
The problem here is more difficult than computing GCDs and has been handled
by O. Ore~\cite{bib-ore} from a theoretical point of view.
We define the two LCMs as
\[
\begin{aligned}
lc=\llcm(a,b)\quad\text{iff}\quad
 &lc\text{ has minimum degree such that}\\
 &\exists A,B,\quad lc=a*A\quad\text{and}\quad lc=b*B;
\end{aligned}
\]
\[
\begin{aligned}
rc=\rlcm(a,b)\quad\text{iff}\quad
 &rc\text{ has minimum degree such that}\\
 &\exists A,B,\quad rc=A*a\quad\text{and}\quad rc=B*b.
\end{aligned}
\]

In order to characterize the preceding \textit{rlcm}, we consider the
following remainder sequence generated by the \textit{rgcd}
applied to two operators \(F _ 1\)  and \(F _ 2 \)
\begin{align*}
      F _ 1 & = Q _ 1 * F _ 2 + F _ 3  \\
      F _ 2 & = Q _ 2 * F _ 3 + F _ 4  \\
     ~ & ...  \\
     ~ & ...  \\
      F _ {n-2} & = Q _ {n-2} * F _ {n-1} + F _ n  \\
      F _ {n-1} & = Q _ {n-1} * F _ n .
\end{align*}

The following result is due to O. Ore~\cite{bib-ore}.

\noindent\textbf{Lemma: }
\textit{To a multiplicative constant the \(\rlcm(F _ 1, F _ 2)\) is}
\[
   F _ {n-1} F _ n ^ {-1} F _ {n-2} F _ {n-1} ^ {-1}
     ....
   F _ 3 F _ 4 ^ {-1} F _ 2 F _ 3 ^ {-1} F _ 1 .
\]
\noindent
We shall not give a proof of this lemma (see~\cite{bib-ore});
instead we concentrate only on the constructive part by giving a symbolic
demonstration.
To begin, define
\begin{align*}
      \delta _ i & = F _ i * F _ {i+1} ^ {-1} \\
      \mu _ i & = F _ {i+1} * F _ i ^ {-1}  .
\end{align*}
Note that \( \delta _ i * \mu _ i = 1\) and especially that
\[
    \delta _ i = (Q _ i F _ {i+1} + F _ {i+2})F _ {i+1}^{-1}
                = Q _ i + \mu _ {i+1}.
\]
\noindent
We also see that \( \delta _{ n-1} = Q _{n-1} \).

Let us build \(
       \delta _ 2 * F _ 1 =(Q _ 2 + \mu _ 3 ) * F _ 1   .
\)
If we define \(\sigma _ 2 = Q _ 2\)
and \(\theta _ 2 = 1\) then the preceding quantity is equal to
\(
       ( \sigma _ 2 + \mu _ 3 * \theta _ 2 ) * F _ 1    .
\)
More generally we have
\(
      \delta _ i \cdots  \delta _ 2 F _ 1 =
                 ( \sigma _ i + \mu _{i+1} \theta _ i ) F _ 1   ,
\)
where
\begin{align*}
      \sigma _{i+1} &= \theta _ i + Q _{i+1} \sigma _ i 
      \\
      \theta _{i+1} &= \sigma _ i  .
\end{align*}
This can be shown using induction:
\begin{align*}
\delta _{i+1}  \cdots  \delta _ 2 F _ 1
& = \delta _{i+1} ( \sigma _ i + \mu _{i+1} \theta _ i ) F _ 1
\\
& = ( ( Q _{i+1} + \mu _{i+2} ) \sigma _ i + \theta _ i ) F _ 1
\\
& = ( \sigma _{i+1} + \mu _{i+2} \theta _{i+1} ) F _ 1   .
\end{align*}
When \(i=n-2\) we have
\begin{align*}
\delta _{n-1} [\sigma _ {n-2} +\mu _{n-1} \theta _{n-2}] F _ 1
   & = (\delta _{n-1} \sigma _ {n-2} + \theta _ {n-2} ) F _ 1
\\
  & = (\sigma _{n-1} + \mu _ n \theta _ n) F _ 1 .
\end{align*}
For the value of \(n\) where \(\mu _ n = 0\),  we have
\(\sigma _{n-1}\) such that
\[
       \rlcm(F _ 1 , F _ 2) = \sigma _{n-1} F _ 1   .
\]

The sequence of values of \(\sigma _ i\) is given by
\begin{align*}
    &\theta _ 1 = 0
                \\
    &\sigma _ 1 = 1
                \\
    &\theta _{i+1} = \sigma _ i
                \\
    &\sigma _{i+1} = \theta _ i + Q _{i+1} \sigma _ i 
                \\
    &\return  {\sigma _{n-1}}
\end{align*}

Now we construct an operator \(\rho_{n-1}\) such that
\(\rlcm(F_1,F_2)=\rho_{n-1}F_2\).
The induction is slightly different from the previous one.

We first build
\[
    F _ 3 ^ {-1} F _ 1
    = F _ 3 ^ {-1} (Q _ 1 F _ 2 + F _ 3 )
    = F _ 3 ^ {-1}  Q _ 1 F _ 2 +1  ,
\]
Then as
\[
    F _ 2 F _ 3 ^ {-1} F _ 1
          = F _ 2 F _ 3 ^ {-1} Q _ 1 F _ 2 + F _ 2
          = (F _ 2 F _ 3 ^ {-1} Q _ 1 + 1) F _ 2    .
\]
we recognize
\begin{align*}
    F _ 2 F _ 3 ^ {-1} F _ 1
        &  = (\delta _ 2 Q _ 1 +1) F _ 2
        \\
       &  = ((Q _ 2 + \mu _ 3)Q _ 1 + 1 )F _ 2
        \\
       &  = ((1 + Q _ 2 Q _ 1) + \mu _ 3 Q _ 1) F _ 2  .
\end{align*}
If we define 
\begin{align*}
       \rho _ 2 & = 1 + Q _ 2 Q _ 1 \\
       \nu _ 2  & = Q _ 1,
\end{align*}
we have
\[
    \delta _ 2 F _ 1 = ( \rho _ 2 + \mu _ 3 \nu _ 2 ) F _ 2  .
\]
It is important to note that
\[
     F _ 1 = (\rho _ 1 + \mu _ 2 \nu _ 1) F _ 2
\]
and \(\rho _ 1 = Q _ 1\) and \( \nu _ 1 = 1\).
By induction it is now easy to prove that we can
build a double sequence of polynomials
\(( \rho _ i,  \nu _ i )\) such that
\begin{align*}
     \rho _{i+1} & = Q _{i+1} \rho _ i + \nu _ i 
      \\
     \nu _{i+1} & = \rho _ i 
\end{align*}
On the iteration when \(\mu _ n = 0\) we have \(\rho _{n-1}\)
as the result.
The sequence of values for \(\rho _ i\) is then
\begin{align*}
     &\rho _ 1 = Q _ 1 
            \\
     &\nu _ 1 = 1 
            \\
     &\rho _{i+1} = Q _{i+1} \rho _ i + \nu _ i 
            \\
     &\nu _{i+1} = \rho _ i 
            \\
     &\return  {\rho _{n-1} } .
\end{align*}

The algorithm for the right LCM expressed in \Language{} is shown in
Figure~\ref{fig-lcm}.
\begin{figure}[t]
\begin{figureframe}
\begin{small}
\begin{code}
llcm(a,b) ==
    a=0 or b=0 => 0
    b0 := b
    u  := 1
    v  := 0
    while lc b ^= 0 repeat
        qr     := ldiv(a,b)
        (a, b) := (b, qr.remainder)
        (u, v) := (u*qr.quotient+v, u)
    b0*u
 
rlcm(a,b) ==
    a=0 or b=0 => 0
    b0 := b
    u  := 1
    v  := 0
    while lc b ^= 0 repeat
        qr     := rdiv(a,b)
        (a, b) := (b, qr.remainder)
        (u, v) := (qr.quotient*u+v, u)
    u*b0
\end{code}

\end{small}
\end{figureframe}
\caption{Left and Right LCM}
\label{fig-lcm}
\end{figure}

Sometimes it is the multipliers \(\sigma _{n-1}\) and \(\rho _{n-1}\)
which are desired.
These can be computed simultaneously as shown in
Figure~\ref{fig-lcmmult}
\begin{figure}[t]
\begin{figureframe}
\begin{small}
\begin{code}
rlmultipliers(a,b) ==
    a = 0  => [1, 0]
    b = 0  => [0, 1]
    a0     := a
    b0     := b
    (s, t) := (0, 1)
    (u, v) := (1, 0)
    while lc b ^= 0 repeat
        qr     := rdiv(a,b)
        (a, b) := (b, qr.remainder)
        (s, t) := (qr.quotient*s+t, s)
        (u, v) := (qr.quotient*u+v, u)
    [s*a0, s, u]  -- s*a0 = u*b0
\end{code}

\end{small}
\end{figureframe}
\caption{Right LCM with Multipliers}
\label{fig-lcmmult}
\end{figure}
\subsection{The Extended Euclidean Algorithm}

We give an extended Euclidean algorithm for a non-commutative ring.
More precisely, we give a constructive proof of the following result:

If \(F1\) and \(F2\) are
linear ordinary differential operators with
coefficients from a field,
then there exist \(A, B, C, D\) such that
\begin{gather*}
{A*{F _ 1} + B*{F _ 2} = \rgcd(F _ 1 ,F _ 2)}
\\
{{F _ 1}*C + {F _ 2}*D = \lgcd(F _ 1 ,F _ 2)}
\end{gather*}

The proof is similar to the commutative case.
We will give the details for the right GCD.
If we take \(A _ 1 = 1 \) and \(B _ 1 = 0 \) it is trivially true
that
\[
         A _ 1 F _ 1 + B _ 1 F _ 2 = F _ 1   .
\]

In the same manner we can take \(A _ 2 = 0\) and \(B _ 2 = 1\)
so that
\[
         A _ 2  F _ 1 + B _ 2 F _ 2 = F _ 2  .
\]
Then by induction let us suppose that there exist two sequences
of polynomials \(A _ i \) and \(B _ i \) such that
\begin{gather*}
     {A _ i F _ 1 + B _ i F _ 2 = F _ {i+1} }
        \\
     {i = 1, \cdots , j . }
\end{gather*}
(where \(F_i\) is generated by the Euclidean algorithm for right GCDs.)
Then \(F _ {j+1} = F _ {j-1} - Q _ {j-1}F _ j\) can be written as
\[
     F _ {j+1} = (A _ {j-1} F _ 1 + B _ {j-1} F _ 2 )
             - Q _ {j-1} (A _ j F _ 1 + B _ j F _ 2 )  .
\]

\noindent
So if we take
\begin{gather*}
     {A _ {i+1} = A _ {i-1} - Q _ {i-1} A _ i }
      \\
     {B _ {i+1} = B _ {i-1} - Q _ {i-1} B _ i }
\end{gather*}
we have the result.
The algorithm is displayed in Figure~\ref{fig-exeucl}.
\begin{figure}[t]
\begin{figureframe}
\begin{small}
\begin{code}
rgcdex(a,b) ==
    a = 0 => [b, 0, 1]
    b = 0 => [a, 1, 0]
    a0 := a
    b0 := b
    (s, t) := (1, 0)
    (u, v) := (0, 1)
    while degree b > 0 repeat
        qr := rdiv(a,b)
        (a, b) := (b, qr.remainder)
        (s, u) := (u, s - qr.quotient*u)
        (t, v) := (v, t - qr.quotient*v)
    if b = 0 then
        [a, s, t]       -- a = s*a0 + t*b0
    else
        [b, u, v]       -- b = u*a0 + v*b0
\end{code}
\end{small}
\end{figureframe}
\caption{Right Extended Euclidean Algorithm.}
\label{fig-exeucl}
\end{figure}
\subsection{Pseudo-Division}

As is easily seen in examples the division process has (at least)
two drawbacks.
The first is that if we start with coefficients in the ring
\(K[x]\) the answer will generally have its coefficients in the field
\(K(x)\).
The second difficulty is a result of the classical ``swell'' of
intermediate results.
In this article we will give only some preliminary results on the first
problem.
We give an algorithm which looks like the ``primitive remainder
sequence'' for the commutative Euclidean division.

It is easily seen that there is no normalization problem for right
division.
Consider the operators
\[
         a := \lc(a) \partial ^ n + \cdots  + a _ 0
\]
and
\[
         b := \lc(b) \partial ^ m + \cdots + b _ 0   ,
\]
where \(n \ge m\).
The main difficulty arises in the following way:

We can begin the pseudo-division process by setting \(r _ 1\) to
satisfy
\[
         \lc(b) a = b \lc(a) \partial ^ {n-m} + r _ 1.
\]
and indeed this gives \(r _ 1\) in \(K[x]\).
However \(b\) does not, in general, divide \(r _ 1 \) on the left.
The usual trick is to multiply both sides by \(\lc(a)\) again.
But we are faced with the product \(\lc(b) ~ b\) and this is
non-commutative.
If we try to commute the two factors then we can destroy
the division process, as in the following example:

Let \(a = x\partial^2+1\) and \(b=x^2\partial+1\).
Then
\[
         x ^ 2 a = b ~ x \partial  + r _ 1   ,
\]
and immediately
\[
        r _ 1 = - x ( x + 1) \partial + x ^ 2  .
\]
If we now form
\begin{align*}
x ^ 2 (x ^ 2 a)
 & = x ^ 2 (x ^ 2 \partial + 1) + x ^ 2 r _ 1
\\
& = b (x ^ 3 \partial)+(-x ^ 4 \partial -x ^ 3 \partial + x ^ 4)
\end{align*}
we cannot continue.

The simplest solution for escaping this (cruel) dilemma is the following.

\textit{In general}, the number of steps of the division process
is \(n-m+1\).
So we begin by multiplying \(a\) by \(\lc(b) ^ {n-m+1}\).
Then we can prove that the first ``remainder'' has \(\lc(b) ^{n-m}\)
as a left factor.

In fact
\[
    \lc(b) ^ {n-m+1} a
          = b \lc(a) \lc(b) ^ {n-m} \partial ^ {n-m} + r
\]
so if we look more deeply at the product in the second member
we have
\[
\begin{aligned}
&\lc(b)\Bigl[\partial^m\bigl(\lc(a)\lc(b)^{n-m}\bigr) +\lc(b)^{n-m}\lc(a)\partial^m\Bigr]\\
&+b_1\Bigl[\partial^{m-1}\bigl(\lc(a)\lc(b)^{n-m}\bigr)
+\lc(b)^{n-m}\lc(a)\partial^{m-1}\Bigr]\\
&+\cdots+b_0\lc(a)\lc(b)^{n-m}.
\end{aligned}
\]
We see that \(r\) can only keep terms with \(\lc(b) ^{n-m}\)
as a factor.
The ones with the power \(n-m+1\) cancel and, more importantly,
\(\partial ^ k (\lc(a) \lc(b) ^ {n-m} )\) is built of two parts ---
a polynomial one, \( \partial ^ k (\lc(a)\lc(b)^ {n-m})\),
which presents no problems for the process and
a ``differential'' one, \(\lc(a)\lc(b) ^ {n-m} \partial ^ k \),
which keeps the right factor.

The program in Figure~\ref{fig-pseudo} implements this procedure.
\begin{figure}[t]
\begin{figureframe}
\begin{small}
\begin{code}
pldiv(a,b) ==
       s: A := lc(b)**(degree(a)-degree(b)+1)
       q: $ := 0
       r: $ := s*a
       while degree r >= degree b and r ^= 0 repeat
           c := lc(r) quo lc(b)
           h := monom(degree(r)-degree(b), c)
           q := q + h
           r := r - b*h
       [q, r]
\end{code}

\end{small}
\end{figureframe}
\caption{Left Pseudo-Division.}
\label{fig-pseudo}
\end{figure}

Two examples of the use of this algorithm follow.

\noindent\textbf{Example 1: }
In this first example we take
\begin{align*}
    a & =17 \partial ^ 8 +1
     \\
    b & =19 x \partial+1 
\end{align*}
and the result of pseudo-division is
\[
\begin{aligned}
\big [\,&15195819563x^7\partial^7-107170516918x^6\partial^6
 +648663655030x^5\partial^5\\
&\quad-3277458467520x^4\partial^4
 +13282331684160x^3\partial^3
 -40546065141120x^2\partial^2\\
&\quad+83226133710720x\partial-87606456537600,\\
&16983563041x^8+87606456537600\,\big].
\end{aligned}
\]
We learn two things from this example.
The first is that the result is (in a certain sense) optimal.
The result has no power of x as a factor.
In other words, the number of steps in the algorithm is \(n-m+1\) (here 8).
The second is that this choice of coefficients
leads to an intermediate expression swell.
In particular we see how important is the fact that
the coefficients were coprime.
It is clear that these kinds of problems need more research.

\noindent\textbf{Example 2: }
In this example we use the two operators
\begin{align*}
   a & =\partial ^ 4 +x ^ 7  \partial ^ 2 + x  \partial + 1 
      \\
   b & =x ^ 3  \partial ^ 2 + \partial + ( x ^ 4 + 34 ) 
\end{align*}
to get
\[
\begin{aligned}
\mathopen{[}\,&x^9\partial^2+(-18x^8-x^6)\partial
 +(x^{16}-x^{10}+216x^7-34x^6+21x^5+x^3),\\[2pt]
&\quad\,(-32x^{18}-x^{16}+x^{13}+38x^{12}+2x^{10}-2016x^9\\
&\qquad{}+1020x^8-252x^7+68x^6-21x^5-x^3)\partial\\
&+(-x^{20}-240x^{17}-34x^{16}-16x^{15}+x^{14}+x^{12}\\
&\qquad{}-126x^{11}-68x^{10}-11x^9-9072x^8\\
&\qquad{}-6325x^7-776x^6-510x^5-111x-34x^3-3x^2)\,\mathclose{]}.
\end{aligned}
\]
In this example we see that the result has \(x ^ 2\) as a factor
so this proposed algorithm is certainly not optimal.
On the other hand we see that results are reasonable.
\subsection{Example: Differential operators with rational function coefficients}

This continues the example from section 1.
We clear the values of most of the variables, but retain those of
\texttt{PZ} and \texttt{PQ}.
\begin{small}
\begin{session}
)clear value Dx a b p e f
 
(Dx, a, b): LODO(QF PZ, QF PZ)
 
\end{session}
\begin{session}
Dx := D()
 
   (16)  "D"
\end{session}
\begin{session}
b := 3*x**2*Dx**2 + 2*Dx + 1/x
 
           2 2         1
   (17)  3x D  + 2D + ---
                       x
\end{session}
\begin{session}
a := b*(5*x*Dx + 7)
 
            3 3       2        2          7
   (18)  15x D  + (51x  + 10x)D  + 29D + ---
                                          x
\end{session}
\begin{session}
p: QF PZ := x**2 + 1/x**2
 
           4
          x  + 1
   (19)  --------
             2
            x
\end{session}

\end{small}
\noindent
Since the operator coefficients depend on \(x\)
the operator multiplication is not commutative.
\begin{small}
\begin{session}
(a*b - b*a) p
 
               4
          - 75x  + 540x - 75
   (20)  --------------------
                   4
                  x
\end{session}

\end{small}
\noindent
Since the coefficients come from a field
it is possible to define left and right division of the operators.
The results of \textit{ldiv} and \textit{rdiv} are
quotient/remainder pairs.
\begin{small}
\begin{session}
ldiv(a,b)
 
   (21)  [5x*D + 7,0]
\end{session}
\begin{session}
-- "%" means the result of the previous expression
 
a - (b * %.quotient + %.remainder)
 
   (22)  0
\end{session}
\begin{session}
rdiv(a,b)
 
                          5
   (23)  [5x*D + 7,10D + ---]
                          x
\end{session}
\begin{session}
a - (%.quotient * b + %.remainder)
 
   (24)  0
\end{session}

\end{small}
\noindent
The divisions allow the computation of left and right
greatest common divisors via remainder
sequences, and consequently the computation of left and right
least common multiples.
\begin{small}
\begin{session}
e := lgcd(a,b)
 
           2 2         1
   (25)  3x D  + 2D + ---
                       x
\end{session}

\end{small}
\noindent
A GCD does not necessarily divide a and b on both sides---here
the left GCD does not divide \(a\) on the right.
\begin{small}
\begin{session}
ldrem(a, e)   -- the remainder from left division
 
   (26)  0
\end{session}
\begin{session}
rdrem(a, e)   -- the remainder from right division
 
                5
   (27)  10D + ---
                x
\end{session}

\end{small}
\noindent
Likewise, an LCM is not necessarily divisible from both sides.
\begin{small}
\begin{session}
f := rlcm(a,b)
 
   (28)
                     4      3             3        2
         5 5     684x  + 80x    4    5832x  + 1656x  + 80x   3
      20x D  + (--------------)D + (-----------------------)D
                      3                         9

                2
           3672x  + 2040x + 352   2     172       - 28
      +  (----------------------)D +  (-----)D + ------
                    9                   9x          2
                                                  9x
\end{session}
\begin{session}
rdrem(f, b)
 
   (29)  0
\end{session}
\begin{session}
ldrem(f, b)
 
           - 1176x + 160       312x - 80
   (30)  (---------------)D + -----------
                9x                  2
                                  9x
\end{session}

\end{small}
\section{The Localization Process in Non-Commutative Algebra}

We want to review the main lines of the classical constructions.
The localization process has its roots in the work of O. Ore~\cite{bib-ore}
in the theory of non-commutative polynomials.
Actually this work has been greatly expanded upon with the work of
P. M. Cohn and his students~\cite{bib-cohn}.
We follow the works of Dixmier, Cohn, and Ore~\cite{bib-dixmier,bib-cohn,bib-ore}.

We start with a ring \(R\) with unity and without zero divisors,
which in general may be non-commutative.
\(R ^ {*}\) will stand for the semigroup of non-zero elements of
\(R\).
Let \(S\) be a subsemigroup of \(R ^ {*}\)
(i.e. \(1 \in S\) and \(S\) is multiplicatively closed).

The goal of this section is to build a ring \(R _ S\) in
which all the elements of \(S\) will have an inverse.
The construction is similar to the commutative one: we build the
Cartesian product \(R\times S\) and we put an equivalence
relation on it
\begin{align*}
(a _ 1, s _ 1) \sim (a _ 2, s _ 2) ~~~~~ \text{iff} ~~~~~
  \exists u, ~ v ~ \in ~ R ~ \text{such that} 
  \\
         s _ 1 * u = s _ 2 * v  ~ \in ~S
  \\
         a _ 1 * u = a _ 2 * v 
\end{align*}
To assure that this relation is an equivalence one it is necessary to
have an extra hypothesis, i.e.

\noindent\textbf{Hypothesis: }
\textit{
We assume that \(S\) is such that
for all \(a\) belonging to \(R ^ {*}\) and
for all \(s\) in \(S\)
we have
\[
       aS \cap sR^{*} \ne \emptyset .
\]
}

A useful lemma is given by Dixmier:

\noindent\textbf{Lemma: }
\textit{
If \((a, s) \sim (a _ 1, s _ 1)\)
and if there exist \(\alpha\) and \( \beta\) in \(R\)
such that \( s * \alpha = s _ 1 * \beta\),
then \( a * \alpha = a _ 1 * \beta \).
}

Using the previous hypothesis and the lemma it is possible to
prove the transitivity of the relation.

We now denote \(B = (R\times S)/\sim \),
the set of equivalence classes, and write \(a/s\) for the class of \((a,s)\).
We are then able to prove the important result~\cite{bib-ore}.

\noindent\textbf{Theorem: }
\textit{
It is possible to put a ring structure on \(B\).
}

We present the constructions for
the main operations of this ring.
\subsection{Addition}

If \(a/s\) and \(b/t\) belong to \(B\) we know from the previous lemma
that there exists a right common denominator,
i.e. \(\alpha\) and \(\beta\) in \(R\) such that
\( s*\alpha =t * \beta = \gamma\).
We also know that we can take \(\alpha\) in \(S\), with the
consequence that \(\gamma\) belongs to \(S\).
So we can define \(a/s+b/t = (a * \alpha + b * \beta )/\gamma\).
To show that this definition is consistent is a rather lengthy task.
\subsection{Multiplication}

Again, let \( a/s \) and \( b/t\) be two elements of \(B\).
As \(s\) belongs to \(S\) and \(b\) to \(R\),
we know that there exists \(u\) in \(S\) and
\(c\) in \(R\) such that \( s * c = b * u \).
In that case we put
\[
          (a/s)(b/t) = (a c)/(t u).
\]
To demonstrate that this definition is consistent, the following lemma is
needed.

\noindent\textbf{Lemma: }
\textit{
If \(c ^\prime \in R \) and \(u ^\prime \in S \) are
such that \( b * u ^\prime = s * c ^\prime \), then
\(
          (a c)/(t u) = (a c^\prime)/(t u^\prime).
\)
}

It is necessary, of course, to verify the ring axioms.
An important point is that the map from \(R\) into \(B\) is an
injective ring homomorphism.
Thus we may identify \(R\) with its image and write its elements as \(a s^{-1}\).

After all these generalities, let us try to come back to earth.
The question to answer is, are there rings which satisfy the main hypothesis?
The answer is certainly positive.
The rings which verify the hypothesis are called Ore rings.
More formally:

\noindent\textbf{Definition: }
\textit{
A ring \(R\) with unity and without zero divisors  is a
\textit{right Ore ring}
if, for all \(a\) and \( b \in R ^ {*} \),
\(aR \cap bR \ne \emptyset\).
}

So, if \(R\) is a right Ore ring, we can choose \( S = R ^ {*} \)
and we can consider \(R_{R^{*}}\), the localization of \(R\)
relative to \(S\).
If it is a ring such that all its elements have an inverse then it is also
a skew field (of course non-commutative).
After Ore we call it the \textit{right field of fractions of \(R\)}.
There are several examples of such rings in~\cite{bib-cohn}.
The only one that we use in this article is the ring of ordinary
differential operators on (say) the ring \(K[x]\).
(The generalisation to other Weyl algebras would be an interesting area
for computer algebra.)

Now we are going to say some words about implementing the localization.
We define the domain constructor \texttt{RightQuotientField} (abbreviated
\texttt{RQF}) to give the localization of \texttt{LODO(A, M)}.
Since \texttt{RQF} requires a ring with a right LCM function,
we make the restriction that \texttt{A} must be a field.
The categorical information of \texttt{RQF} is
\begin{small}
\begin{code}
RightQuotientField: (A, M) -> RQFCAT where
 
    A: DifferentialField
    M: Module(A) with deriv: $ -> $
 
    RQFCAT == Join(SkewField, Module(LODO(A, M)) with
        numer: $ -> LODO(A, M)
        denom: $ -> LODO(A, M)
        ".":   ($, M) -> M
\end{code}

\end{small}
\noindent
Thus the operations of \texttt{RQF(A, M)} include
\begin{small}
\begin{code}
"=":    ($,$) -> Boolean                 ".":  ($, M) -> M
0:      () -> $                          "-":  $ -> $
1:      () -> $                          "-":  ($,$) -> $
"+":    ($,$) -> $                       "*":  ($,$) -> $
"*":    (I,$) -> $                       "*": (LODO(A, M), $) -> $
"/":    (LODO(A, M), LODO(A, M)) -> $    "**":   ($,NonNegativeInteger) -> $
denom:  $ -> LODO(A, M)                  numer:  $ -> LODO(A, M)
recip:  $ -> Union($,failed)             coerce: I -> $
coerce: LODO(A, M) -> $                  coerce: $ -> Expression
characteristic : () -> NonNegativeInteger
\end{code}
\end{small}
We present in Figure~\ref{fig-local}
the functions for addition and multiplication.
\begin{figure}[t]
\begin{figureframe}
\begin{small}
\begin{code}
-- rightMultipliers(u,v) gives [w,p,q], where w = u*p = v*q.
rightMultipliers(a,b) ==
       a = 0 => [0, 1, 0]
       b = 0 => [0, 0, 1]
       a0 := a
       b0 := b
       (s, t) := (0, 1)
       (u, v) := (1, 0)
       while lc b ^= 0 repeat
           qr := ldiv(a,b)
           (a, b) := (b, qr.remainder)
           (s, t) := (s*qr.quotient+t, s)
           (u, v) := (u*qr.quotient+v, u)
       [a0*s, s, u]  -- a0*s = b0*u
 
a, b: $           -- $ is RQF(A, M)
n, d: LODO(A, M)
 
a + b ==
       l := rightMultipliers(denom a, denom b)
       d := l.1
       n := numer a * l.2 + numer b * l.3
       n / d
 
a * b ==
       l := rightMultipliers(denom a, numer b)
       n := numer a * l.2
       d := denom b * l.3
       n / d
\end{code}

\end{small}
\end{figureframe}
\caption{Operations in the Localization.}
\label{fig-local}
\end{figure}
\section{Factorization in a Non-Commutative Domain}

The problem we want to touch upon is the following:
\textit{Given an ordinary differential operator can we split it into factors
belonging to the same ring?}
That is, if L belongs to \(K(x)[\partial]\), we want to
factor so that the factors also belong to \(K(x)[\partial]\).
The problem has been studied by a number of people
(Frobenius~\cite{bib-frobenius}, Picard~\cite{bib-picard}, and Valiron~\cite{bib-valiron}), though generally for
second order operators.

It is important to note before we start that the factorization in
\(K(x)[\partial]\)
is not necessarily unique.
A classic counterexample is
\[
\partial^2-\frac{2}{x+1}\partial+\frac{2}{(x+1)^2}=
\]
\[
\left(\partial-\frac{1}{\omega(x+1)^2+x+1}\right)
*\left(\partial-
\frac{2\omega(x+1)+1}{\omega(x+1)^2+x+1}\right),
\]
with equality for all \(\omega\).

In this first paper we also consider this case ---
first because it is interesting in its own right and secondly
because it is a powerful model for the general case of linear
operators.
This is a generalization of the approach of Rainville~\cite{bib-rainville}
implemented in Macsyma~\cite{bib-lafferty}, where factors in \(K[x][\partial]\)
are sought.
\subsection{The Problem}

Let \(L=\partial ^ 2 + a _ 1 \partial + a _ 0 \) be an element
of \(K(x)[\partial]\).
The classic approach is to try a factorisation in the form
\[
      L=( \partial + \alpha ) * ( \partial + \beta ).
\]
As we are looking for a factorization over \(K(x)\) we can
suppose that
\(\alpha\) and \(\beta \) belong to this field.

Expanding the product for \(L\) we obtain
\[
      L = \partial ^ 2 + ( \alpha + \beta ) \partial
                        + ( \alpha \beta +\beta ^\prime ).
\]
and identifying the coefficients yields
\begin{gather*}
      \alpha +\beta = a _ 1
       \\
      \alpha \beta + \beta ^\prime = a _ 0.
\end{gather*}
We see that \(\beta \) must satisfy the special Riccati equation
\[
       \beta ^\prime = \beta ^ 2 - a _ 1 \beta + a _ 0
\]
Of course this equation is a non-linear one but
\textit{the important point is that we are looking for rational solutions
of this equation.}

Though presented in a slightly different fashion, this is the starting
point of Picard's analysis of the problem.

The key to the algorithm is a careful examination of the possible
poles of the solution.
These poles can come from two different sources:

1) from singularities of the Riccati equation
--- these singularities
being located at the poles of \(a _ 1\) and \(a _ 0\) and
at infinity.

2) From zeros of the solutions to the linear equation \(L = 0\).
In fact
it is easily seen that \(\beta\) is nothing more than the logarithmic
derivative of a solution of \(L\) so the possible singularities
are the poles of \(\beta \).

So, if \(\beta\) exists, it has the following form
\[
   \beta = \sum _{ i=1}^m
        \left (
           \sum _{j=1}^{n _ i}
              \frac{A^i_j}{(x-\alpha_i)^j}
        \right )
        + \theta _ 0 + \cdots + \theta _ m x ^ m
        + \gamma
\]
In this formula the first sum is associated with the singularities of
\(a _ 0\) and \(a _ 1 \).
\( \gamma \) is the rational function associated with the zeros of
\(L\) and the polynomial with coefficients
\( \theta _ i\) is associated with
the behavior  of \( \beta \) at infinity.
\subsection{The rational part associated with the solution of L}

We now study the rational part of \(\beta\) associated with
the zeros of the solution of \(L\).
One approach is to look for the order of the poles of
\(\gamma\).
Let us suppose that one of these poles is zero.
Then, as zero is a regular point of the solution, zero will not be a
singular point of \(a _ 0\) nor of \(a _ 1\).

For \(\gamma\) we can write
\[
     \gamma = \frac{\tau}{x^p} + \cdots .
\]
The other terms are of order less than \(p\).
Since this must satisfy the Riccati differential equation, we obtain
\[
-\frac{p\tau_p}{x^{p+1}} + \cdots =
\]
\[
               \left (
                    \frac{\tau_p^2}{x^{2p}} + \cdots
               \right )
               + a _ 1
                    \left (
                         \frac{\tau_p}{x^p} + \cdots
                    \right )
               + a _ 0
\]
Because \(a _ 0\) and \(a _ 1 \) do not have poles at zero, we
see that the solution is \(p+1\) = \( 2 p \) and also
\( \tau _ p = -1\).
Now we can say that
\[
   \beta = -\left(
                \frac{1}{x-\alpha_0}
                    + \cdots +
                \frac{1}{x-\alpha_\nu}
            \right).
\]
If we introduce
\[
    \sigma (x) = \prod_{i=0}^\nu {( x - \alpha _ i )}
\]
\noindent
then
\[
     \gamma = -\frac{\sigma'(x)}{\sigma(x)}.
\]

We actually have the form of the solution and what we need now is to
identify the coefficients.
\subsection{Coefficient Identification}

The first method (Picard's) is to determine the coefficients
\(A ^ i _ j \) and \( \theta _ i \).
The idea is to translate the operator at the roots of the denominators of
\(a _ 1\) and \(a _ 0 \).
In order to separate the problems let us suppose that one of these roots is
located at zero.

We can write the Riccati equation in the following form
\[
       x ^ \nu \beta ^\prime = x ^ \nu \beta ^ 2
                             + b _ 1 \beta
                             + b _ 0 .
\]
Here we make the hypothesis that \(b _ 1\) and \( b _ 0 \)
are of the form
\(b _ 0 =   x ^ \rho a _ 0\) and
\(b _ 1 = - x ^ \mu  a _ 1\),
where \(\rho\) and \(\mu\) are non-negative integers and are not
both zero.

To be able to study easily the possibility of a pole at zero, we
want to introduce a Newton polygon for our equation.
This polygon is inspired by Ince~\cite{bib-ince} and by related work of Della Dora and Watt~\cite{bib-delladora}.
\subsection{Construction of the Newton Polygon}

A Newton polygon is determined by a number of points in the plane, each
associated with a monomial of the differential equation.
With the differential monomial \(x ^ a \partial ^ b\),
we associate the point \(( b , a-b )\)
in a rectangular coordinate system.
With the monomial \(x ^ a \beta ^ b \) we associate
\((b, a)\) in the same coordinate system.
Then we take the lower convex envelope of these points.

\begin{figure}[t]
\begin{figureframe}
\centering
\begin{tikzpicture}[x=1.6cm,y=1.2cm]
  \draw[->] (-0.2,0) -- (2.5,0) node[below] {$b$};
  \draw[->] (0,-0.2) -- (0,2.5) node[left,align=center] {$a$\\or $a-b$};
  \foreach \x in {0,1,2} {
    \draw (\x,0.06) -- (\x,-0.06) node[below=4pt] {$\x$};
  }
  \foreach \y in {0,1,2} {
    \draw (0.06,\y) -- (-0.06,\y) node[left=4pt] {$\y$};
  }
  \draw[thick] (0,0) -- (2,1);
  \fill (0,0) circle (1.6pt) node[above left=2pt] {$33$};
  \fill (1,1) circle (1.6pt) node[above left] {$x^2\beta^\prime,\ x\beta$};
  \fill (1,2) circle (1.6pt) node[above right] {$2x^2\beta$};
  \fill (2,1) circle (1.6pt) node[above right] {$x\beta^2$};
\end{tikzpicture}
\end{figureframe}
\caption{The Newton polygon for $x^2\beta' = x\beta^2 + x(1+2x)\beta + 33$.}
\label{fig-newton}
\end{figure}
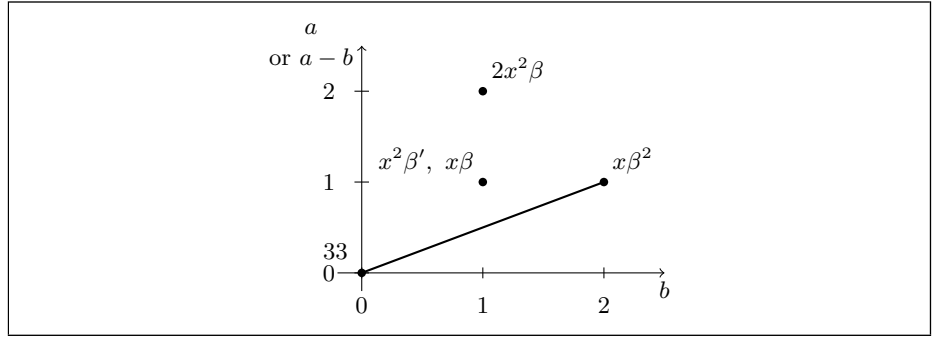

\noindent\textbf{Example: }
Let us consider the following differential equation
\[
     x ^ 2 \beta ^\prime = x \beta ^ 2 + x ( 1 + 2 x ) \beta + 33.
\]
The Newton polygon is shown in Figure~\ref{fig-newton}.
The monomials \(x^2\beta'\) and \(x\beta\) both determine the point
\((1,1)\).

We can prove the following lemma which is the key to the construction
of the solution.

\noindent\textbf{Lemma: }
\textit{
If \(p\) is a slope of the Newton polygon, then \(\beta\) will have
an expansion about zero of the following form:
\[
     \beta(x) = \frac{A_p}{x^p}
             + \cdots
             + \frac{A_1}{x}
             +  \hat \beta ( x )
\]
where \(\hat \beta \) is a holomorphic function at zero.
}

\noindent\textbf{Proof: }
Let \(\ell\) be the line containing the side in question.
The demonstration is based on the simple observation that the
points of the diagram cannot lie on both sides of \(\ell\).
We have three cases:

1) Suppose \(x ^ a \beta ^ b\) is on \(\ell\)
and we have a point from \(x ^ c \beta ^ d\) in the diagram.
Since \(\ell\) is part of the lower convex envelope and
since the point \(x ^ c \beta ^ d\) lies on
a line parallel to \(\ell\),
we have
\(a = p b + q\) and \(c = p d + q ^\prime\) with \(q ^\prime \ge q\).
When we substitute in
\[
      \frac{A}{x^p} + \phi
\]
we have
\begin{gather*}
     x ^ a \beta ^ b \rightarrow \frac{1}{x^{pb-a}}
      \\
     x ^ c \beta ^ d \rightarrow \frac{1}{x^{pd-c}}
\end{gather*}
Now \(q ^\prime > q\) implies
\(
    ~ c-p d > a- p b ~
\)
which is equivalent to
\(
    ~ p b - a > p d - c ~.
\)
Therefore the point with highest polar order lies on \(\ell\).
(Of course there are at least two of them on this line.)

2) Again suppose \(~ x ^ a \beta ^ b ~\) is on \(\ell\)
but this time we have a point from
\(~ x ^ c \partial ^ d ~\) in the diagram.
If we remember that \(d = 1\) then \( c - 1 = p + q ^\prime\)
with \( q ^\prime > q \).
The same argument as above leads to the result.

3) The last part of the argument is to consider that
\(x ^ a \partial ^ b \) is on \(\ell\) and that a point from
\(x ^ c \beta ^ b \) is in the diagram.
We leave the end of the proof to the proverbial lively reader.

\noindent\textbf{Remarks: }
We remark that the points on the edge give the \textbf{characteristic
equation} associated with zero.
It is interesting to note that if \(p\) is not an integer
we have an irreducibility criterion.

Using the previous lemma we are able to determine the order of the pole
at a singularity.
At this point we can put
\[
     \beta = \frac{A_p}{x^p}
            + \cdots
            + \frac{A_1}{x}
            + \hat \beta 
\]
in the Riccati equation, and a method of indeterminate coefficients
gives the coefficients.
We can use the same method at each singular point and at the
infinite point.

\noindent\textbf{Remark: }
It should be clear that this kind of reasoning overcomes two of the
most important difficulties encountered in using computer
algebra in these kinds of problems.

1) As is easily seen, \(A _ p\) may be a root of an algebraic
equation (here of degree less than or equal to 2).
So it may be necessary to work in some composite extension field.

2) As a consequence of \#1 it may be necessary to ``follow'' each root
of the equation.
That is, when we calculate another expansion at another singular
point,
we have to keep track of the several previous expansions and
separate them.

These two points are, of course, absent from Picard's analysis
of the problem.

At this point it is clear that, at least theoretically, the only
thing we do not know is the logarithmic derivative associated with the
zeros of the solutions, i.e. the \(\gamma\) term,
and it is that which gives us the last criterion for
irreducibility.
The method is to make a change of the variable
in the Riccati differential equation
in such a way that the unknown function will be \( \gamma \).

Let us call \(A\) the part of the solution associated with the
singularities other than those of \(\gamma\).
\[
       (A+\gamma) ^\prime = (A+\gamma) ^ 2
                          - a _ 1 ( A  + \gamma )
                          + a _ 0
\]

As we know \(A\), we obtain a new equation
\[
        \gamma ^\prime = \gamma ^ 2 + c _ 1 \gamma + c _ 0.
\]
In this equation \(c _ 1 \) and \( c _ 0 \) do not have
any singularities at the roots of the denominators of \(\gamma\).

Now we try to determine a polynomial \(\sigma\) of degree \(m\)
such that the previous equation will be satisfied.
If we cannot determine  \(m\) by substitution and
identification of the highest degree term, then the equation will be
irreducible.  Otherwise we can construct the polynomial and find the
factors of \(L\).
\subsection{Improvements to the Picard Algorithm}

It is clear that the Picard algorithm must be improved in several  ways.
One of these is relatively easy:
The introduction of the Newton polygon gives us the ability to quickly
analyze the nature of the strength of a singularity at a point.
We can perform the following analysis.

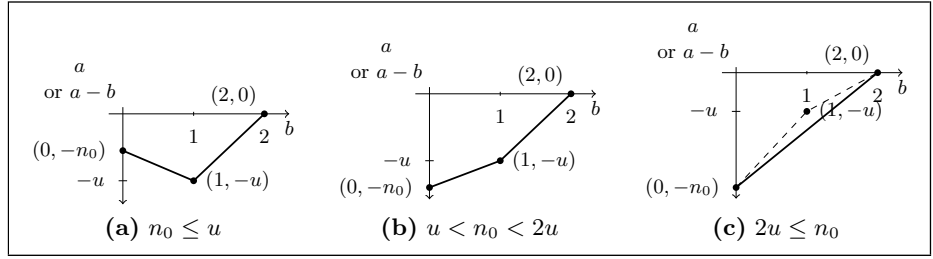
\begin{figure}[t]
\begin{figureframe}
\centering
\begin{minipage}[t]{0.31\linewidth}
\centering
\resizebox{\linewidth}{!}{%
\begin{tikzpicture}[x=1.10cm,y=0.52cm]
  \draw[->] (-0.2,0) -- (2.35,0) node[below] {$b$};
  \draw[->] (0,0.2) -- (0,-2.7);
  \node[above left,align=center] at (0,0.2) {$a$\\or $a-b$};
  \draw (2,0.06)--(2,-0.06) node[below=3pt] {$2$};
  \draw (1,0.06)--(1,-0.06) node[below=3pt] {$1$};
  \draw (0.06,-2)--(-0.06,-2) node[left=3pt] {$-u$};
  \fill (0,-1.1) circle (1.5pt) node[left=4pt] {$(0,-n_0)$};
  \fill (1,-2) circle (1.5pt) node[right=2pt] {$(1,-u)$};
  \fill (2,0) circle (1.5pt) node[above left] {$(2,0)$};
  \draw[thick] (0,-1.1) -- (1,-2) -- (2,0);
\end{tikzpicture}%
}

\textbf{(a)} $n_0\le u$
\end{minipage}\hfill
\begin{minipage}[t]{0.31\linewidth}
\centering
\resizebox{\linewidth}{!}{%
\begin{tikzpicture}[x=1.10cm,y=0.52cm]
  \draw[->] (-0.2,0) -- (2.35,0) node[below] {$b$};
  \draw[->] (0,0.2) -- (0,-3.15);
  \node[above left,align=center] at (0,0.2) {$a$\\or $a-b$};
  \draw (2,0.06)--(2,-0.06) node[below=3pt] {$2$};
  \draw (1,0.06)--(1,-0.06) node[below=3pt] {$1$};
  \draw (0.06,-2)--(-0.06,-2) node[left=3pt] {$-u$};
  \fill (0,-2.8) circle (1.5pt) node[left=4pt] {$(0,-n_0)$};
  \fill (1,-2) circle (1.5pt) node[right=2pt] {$(1,-u)$};
  \fill (2,0) circle (1.5pt) node[above left] {$(2,0)$};
  \draw[thick] (0,-2.8) -- (1,-2) -- (2,0);
\end{tikzpicture}%
}

\textbf{(b)} $u<n_0<2u$
\end{minipage}\hfill
\begin{minipage}[t]{0.31\linewidth}
\centering
\resizebox{\linewidth}{!}{%
\begin{tikzpicture}[x=1.10cm,y=0.43cm]
  \draw[->] (-0.2,0) -- (2.35,0) node[below] {$b$};
  \draw[->] (0,0.2) -- (0,-4.55);
  \node[above left,align=center] at (0,0.2) {$a$\\or $a-b$};
  \draw (2,0.06)--(2,-0.06) node[below=3pt] {$2$};
  \draw (1,0.06)--(1,-0.06) node[below=3pt] {$1$};
  \draw (0.06,-1.4)--(-0.06,-1.4) node[left=3pt] {$-u$};
  \fill (0,-4.15) circle (1.5pt) node[left=4pt] {$(0,-n_0)$};
  \fill (1,-1.4) circle (1.5pt) node[right=2pt] {$(1,-u)$};
  \fill (2,0) circle (1.5pt) node[above left] {$(2,0)$};
  \draw[thick] (0,-4.15) -- (2,0);
  \draw[dashed] (0,-4.15) -- (1,-1.4) -- (2,0);
\end{tikzpicture}%
}

\textbf{(c)} $2u\le n_0$
\end{minipage}
\end{figureframe}
\caption{The three Newton-polygon configurations.  In (a), only the edge
from $(1,-u)$ to $(2,0)$ has positive slope.  In (b), both lower edges
have positive slope and give two candidate pole orders.  In (c), the
single relevant lower edge joins $(0,-n_0)$ to $(2,0)$ and has slope
$n_0/2$.}
\label{fig-cases}
\end{figure}

First we introduce the following notation (under the hypothesis that
the studied singularity is located at zero)
\begin{align*}
    {n _ 1} &= {\text{polar order of}}~ { a _ 1}~{\text{at zero}}
      \\
    {n _ 0} &= {\text{polar order of}}~ {a _ 0}~{\text{at zero}}
\end{align*}
and
\[
      u = \max(1,n _ 1).
\]
Then we have the following three cases:

1) If \(n _ 0 \le u \), then as can be seen in
Figure~\ref{fig-cases}(a), we
only have one solution of the form
\[
      y = \frac{A_u}{x^u}
           + \cdots
           + \frac{A_1}{x}
           + \cdots .
\]

2) If \( u < n _ 0 < 2 u\) then we have two
solutions
\begin{align*}
           y &= \frac{A_u}{x^u}
                 + \cdots
                 + \frac{A_1}{x} + \cdots 
     \\
           y &= \frac{B_{n_0-u}}{x^{n_0-u}}
                 + \cdots
                 + \frac{B_1}{x} .
\end{align*}
This can be seen in Figure~\ref{fig-cases}(b).

3) The last case is \(2 u \le n _ 0 \).
Here we have to distinguish the following two cases:
\begin{itemize}
\item
If \({n _ 0} / 2 \) is not an integer, then there is no solution.
\item
If \({n _ 0} / 2 \) is an integer, then we have two
solutions of the form
\[
          y = \frac{A^{1/2}}{x^{n_0/2}}
              + \cdots
\]
\end{itemize}
This can be seen in Figure~\ref{fig-cases}(c).
\subsection{Other Improvements}

One of the drawbacks of Picard's method is that
we must use full factorization of the various denominators.
Instead we can propose the following scenario.

Let \(d _ 0\) be the denominator of \(a _ 0\)
and let \(d _ 1\) be that of \(a _ 1\).
\par\noindent
\par\noindent We can perform a prime factor decomposition of \(d _ 0\) and
\(d _ 1 \).
\begin{align*}
     {d _ 0} &= u _ 1 ^ {n _ 1} \cdots u _ k ^{n _ k}
  \\
     {d _ 1} &= v _ 1 ^ {m _ 1} \cdots v _ \ell ^{m _ \ell}
\end{align*}

If \(w\) is a factor of one of the previous two polynomials then a root
of \(w\) gives rise to a translation and eventually to a pole of the
solution \(\beta\).

If we write
\[
     \beta ^\prime = \beta ^ 2
                  - \frac{\gamma}{w^p} ~ \beta
                  + \frac{\delta}{w^q}
\]
then calling \(\xi\) one of the roots of \(w\), the translation
\(t = x - \xi\) turns the previous equation into the following
\[
     \beta ^\prime = \beta ^ 2
                - \frac{\gamma(t+\xi)}{t^p\tau^p}~ \beta
                + \frac{\delta(t+\xi)}{t^q\tau^q}.
\]

As we can take \( \gamma\) and \(w\) co-prime, the index
\(p\) will then only depend on \(w\) (identically for \(q\)) and
so the order of the pole is in fact the power of the factor \(w\).
So the decomposition of \(\beta\) for the singularities coming
from \(a _ 0\) and \(a _ 1 \) uses only a prime factor
decomposition.

In the same way, it is clear, that the numerators of the different
factors of the denominators must belong to the previous field.
In other words, no algebraic extension is needed for this part of the
problem.
This may overcome some of the main drawbacks in Picard's presentation.
\subsection{Example}

Let us look at the following simple example:
\[
      L  = \partial ^ 2
           + \frac{x^3+x^2+x+2}{x(x+1)(x+2)} ~ \partial
           + \frac{3x+4}{(x+1)(x+2)^2}    .
\]

The associated Riccati equation is
\[
    {\beta ^\prime} = {\beta ^ 2}
                 - \frac{x^3+x^2+x+2}{x(x+1)(x+2)} ~ \beta
                 + \frac{3x+4}{(x+1)(x+2)^2}  .
\]
If we look at the point \(0\), we observe from the Newton diagram that
the characteristic equation is
\[
      - a = a ^ 2 - a.
\]
\noindent Hence \(a=0\): there is no nonzero characteristic root, and
therefore no pole of the assumed form.
We can, however, observe the important fact that \(x\) appears in the
denominator of \( a _ 1 \) but not in that of \(a _ 0 \).
A closer look shows that this means that the solution must be
of the form:
\[
       \left(\partial + \frac{u}{xv}\right)
       \left(\partial + \frac{wx}{t}\right)
\]

The second important point is \(- 2\).
The Newton polygon tells us that the characteristic equation has two real
roots, \(a=-1\) and \(a=- 2\),
so we can try a solution of the form
\[
      \beta = -\frac{2}{x+2} + \theta.
\]
After this change of variable the equation becomes
\begin{align*}
     \theta ^\prime &= \theta ^ 2
                   - \frac{x^3+5x^2+5x+2}{x(x+1)(x+2)} ~ \theta 
                   \\
                   & + \frac{2x^3+7x^2+8x+4}{x(x+1)(x+2)^2} = 0.
\end{align*}
The problem now is to look at the last singularity, \(-1\).

We learn from the Newton polygon that the solution must be of the form
\[
      \frac{a}{x+1}.
\]
The characteristic equation is \(a^2=0\),
so we do not have any contribution to the solution for the \(-1\)
singularity.

Let us now look at the behavior of our function at infinity.
For the same reason as before we can see that the only possibility
is that the polynomial reduces to a constant.
Immediately this constant can be seen to be equal to \(-1\).
And we see that the constant \(-1\) is a solution of the previous equation.
So we have found the solution
\[
       \beta = \frac{x}{x+2}.
\]
This gives the flavour of the Picard algorithm.

\clearpage
\appendix
\section{Constructors for LODO}

The right algorithms for division, gcd, lcm, etc. are defined in
terms of the left algorithms on the opposite ring.  Therefore only
the left algorithms are shown here.  The complete LODO constructor is
given first; its two supporting constructors are shown afterward in
Figures~\ref{fig-opr} and~\ref{fig-ncdiv}.

\begin{small}
\begin{code}
 
--  This domain defines a ring of differential operators which act upon
--  an A-module, where A is a differential ring.
 
LinearOrdinaryDifferentialOperator(A, M): DOcategory == DOcapsule where
    NNI ==> NonNegativeInteger
    SUP ==> SparseUnivariatePolynomial
 
    A:  DifferentialRing
    M:  Module(A) with deriv: $ -> $
 
    DOcategory == GeneralPolynomialWithoutCommutativity(A,NNI) with
        D:       () -> $
        elt:     ($, M) -> M
 
        coerce:  $ -> SUP(A)
        coerce:  SUP(A) -> $
 
        if A has commutative("*") and A has constant(deriv) then
            commutative("*")
 
        if A has Field then
            ldiv:   ($, $) -> Record(quotient: $, remainder: $)
            ldquo:  ($, $) -> $
            ldrem:  ($, $) -> $
            ldexquo:($, $) -> Union($, "failed")
 
            rdiv:   ($, $) -> Record(quotient: $, remainder: $)
            rdquo:  ($, $) -> $
            rdrem:  ($, $) -> $
            rdexquo:($, $) -> Union($, "failed")
 
            lgcd:   ($, $) -> $
            llcm:   ($, $) -> $
            rgcd:   ($, $) -> $
            rlcm:   ($, $) -> $
 
    DOcapsule == SUP(A) add
        Rep := SUP(A)
 
        coerce(a:$):SUP(A)     == a::Rep::SUP(A)
        coerce(p:SUP(A)):$     == p::Rep::$
 
        coerce(a:$):Expression == output(a:Rep, "D"::Expression)$Rep
 
        D() == monom(1,1)
 
        elt(a, u) ==
            u, w, uderiv: M
            w := 0
            for i in 0..degree a repeat
                uderiv := if i = 0 then u else deriv uderiv
                w := w + coef(a,i::NNI)*uderiv
            w
 
        a:$ * b:$ ==
            alpha, beta, bderiv: A
            aa, bb, r: $
            degaa, degbb, n: NNI
            nCi : Integer
 
            r  := 0
            aa := a
            while aa^=0 repeat
                degaa := degree aa; alpha := lc aa; aa := red aa
                bb := b
                while bb^=0 repeat
                    degbb := degree bb; beta := lc bb; bb := red bb
                    for i in 0..degaa repeat
                        if i=0 then
                            bderiv := beta   -- i-th derivative of beta
                            nCi    := 1      -- degaa choose i
                        else
                            bderiv := deriv bderiv
                            nCi    := nCi * (degaa - i + 1) quo i
                        d := (degaa + degbb - i):NNI
                        r := r + monom(d, nCi*alpha*bderiv)
            r
 
        if A has Field then
            -- The opposite ring of LODO
            Op ==> OppositePolynomialRing($, A, NNI)
 
            -- Define the right algorithms in terms of
            -- the left algorithms on the opposite ring.
            DOdiv := NonCommutativePolynomialDivision($,  A, NNI)
            OPdiv := NonCommutativePolynomialDivision(Op, A, NNI)
 
            -- [q,r] = ldiv(a,b) means a=b*q+r
            -- [q,r] = rdiv(a,b) means a=q*b+r
            ldiv(a, b) == ldiv(a, b)$DOdiv
            rdiv(a,b) ==
                qr := ldiv(op a, op b)$OPdiv
                [po qr.quotient, po qr.remainder]
 
            -- ldquo(a,b) is the quotient from left division, etc.
            ldquo(a,b)   == ldiv.(a,b).quotient
            ldrem(a,b)   == ldiv.(a,b).remainder
            ldexquo(a,b) ==
                 qr := ldiv(a,b)
                 if qr.remainder = 0 then qr.quotient else "failed"
            rdquo(a,b)   == rdiv.(a,b).quotient
            rdrem(a,b)   == rdiv.(a,b).remainder
            rdexquo(a,b) ==
                 qr := rdiv(a,b)
                 if qr.remainder = 0 then qr.quotient else "failed"
 
            -- l = lgcd(a,b) means  a = l*aa  b = l*bb.  Uses ldiv.
            -- l = llcm(a,b) means  l = a*aa  l = b*bb   Uses ldiv.
            lgcd(a,b) == lgcd(a, b)$DOdiv
            llcm(a,b) == llcm(a, b)$DOdiv
 
            -- r = rgcd(a,b) means  a = aa*r  b = bb*r.  Uses rdiv.
            -- r = rlcm(a,b) means  r = aa*a  r = bb*b.  Uses rdiv.
            rgcd(a,b) == po lgcd(op a, op b)$OPdiv
            rlcm(a,b) == po llcm(op a, op b)$OPdiv
\end{code}

\end{small}

\begin{figure}[h]
\begin{figureframe}
\begin{small}
\begin{code}
OppositePolynomialRing(P, R, E): OPRcat == OPRdef where
   P: GeneralPolynomialWithoutCommutativity(R, E)
   R: Ring
   E: OrderedAbelianMonoid
 
   OPRcat == GeneralPolynomialWithoutCommutativity(R, E) with
        if P has DifferentialRing then DifferentialRing
        op: P -> $
        po: $ -> P
 
   OPRdef  == P add
        Rep := P
        x, y: $
        a: P
        op a == a: $
        po x == x: P
        x*y == (y:P) *$P (x:P)
        coerce(x): Expression == mkUnary(op, coerce(x:P)$P)
\end{code}

\end{small}
\end{figureframe}
\caption{Constructor for Opposite Polynomial Ring}
\label{fig-opr}
\end{figure}
\clearpage
\begin{figure}[t]
\begin{figureframe}
\begin{small}
\begin{code}
NonCommutativePolynomialDivision(P, F, E): PDcat == PDdef  where
    P: GeneralPolynomialWithoutCommutativity(F, E)
    F: Field
    E: OrderedAbelianMonoid
 
    PDcat == with
        ldiv:   (P, P) -> Record(quotient: P, remainder: P)
        ldquo:  (P, P) -> P
        ldrem:  (P, P) -> P
        ldexquo:(P, P) -> Union(P, "failed")
 
        lgcd:   (P, P) -> P
        llcm:   (P, P) -> P
    PDdef == add
        ldiv(a, b) ==
            q: P := 0
            r: P := a
            iv:F := inv lc b
            while degree r >= degree b and r ^= 0 repeat
                h := monom((degree r - degree b)::E, iv*lc r)$P
                r := r - b*h
                q := q + h
            [q,r]
 
        -- ldquo(a,b) is the quotient from left division, etc.
        ldquo(a,b)   == ldiv(a,b).quotient
        ldrem(a,b)   == ldiv(a,b).remainder
        ldexquo(a,b) ==
             qr := ldiv(a,b)
             if qr.remainder = 0 then qr.quotient else "failed"
        -- l = lgcd(a,b) means  a = l*aa  b = l*bb.  Uses ldiv.
        lgcd(a,b) ==
             a = 0 =>b
             b = 0 =>a
             while degree b > 0 repeat (a,b) := (b, ldrem(a,b))
             if b=0 then a else b
        -- l = llcm(a,b) means  l = a*aa  l = b*bb   Uses ldiv.
        llcm(a,b) ==
            a = 0 =>b
            b = 0 =>a
            b0 := b
            u  := monom(0,1)$P
            v  := 0
            while lc b ^= 0 repeat
                qr     := ldiv(a,b)
                (a, b) := (b, qr.remainder)
                (u, v) := (u*qr.quotient+v, u)
            b0*u
\end{code}

\end{small}
\end{figureframe}
\caption{Constructor for Noncommutative Polynomial Division}
\label{fig-ncdiv}
\end{figure}

\end{document}